\documentclass[12pt]{article}
\usepackage{cite}
\def\cE{\mathcal{E}}
\usepackage{epsfig}
\usepackage{multicol}
\usepackage[left=1.5cm,top=1.5cm,right=1.5cm,bottom=2cm]{geometry}
\usepackage[small]{titlesec}

\usepackage{amsfonts}
\usepackage{amsmath}
\usepackage{amssymb}
\usepackage{graphicx}
\usepackage{latexsym}
\usepackage{amsmath,dsfont}
\usepackage[latin1]{inputenc}
\usepackage[english]{babel}

\usepackage{appendix}
\usepackage{makeidx}
\usepackage{url}
\usepackage{pdfsync}
\usepackage{color}
\usepackage[
pdfkeywords={},%
pdfsubject={},%
pdftitle={},%
pdfauthor={}%
]{hyperref}
\usepackage{epstopdf}
\usepackage{booktabs}
\usepackage{multirow}

\makeatletter

\newcommand{\half}{{\frac{1}{2}}}

\newcommand{\W}{\Omega}

\newcommand{\be}{\begin{equation}}
\newcommand{\ee}{\end{equation}}
\newcommand{\bea}{\begin{eqnarray}}
\newcommand{\eea}{\end{eqnarray}}

\def\cH{{\mathcal{H}}}
\def\cP{{\mathcal{P}}}

\def\A{\mathbb A}

\def\R{\mathbb R}
\def\W{\mathbb W}

\usepackage{amsmath,amssymb}
\def\barray{\begin{array}}
	\def\earray{\end{array}}

\def\sn{\mathrm{sn}}
\def\cn{\mathrm{cn}}
\def\dn{\mathrm{dn}}

\def\cP{\mathcal{P}}

\def\cS{\mathcal{S}}
\def\cQ{\mathcal{Q}}

\makeatother
\def\R{\mathbb R} \def\I{\mathbb I}

\def\be{\begin{equation}}
\def\ee{\end{equation}}\def\f{\frac}

\def\bea{\begin{eqnarray}}\def\eea{\end{eqnarray}}

\def\ben{\begin{displaymath}}
\def\ba{\begin{array}{c}}\def\bal{\begin{array}{l}}\def\ea{\end{array}}
\def\sn{\mathrm{sn}}\def\cn{\mathrm{cn}}
\def\dn{\mathrm{dn}}

\def\A{\mathbb A}

\def\R{\mathbb R}
\def\cH{{\mathcal{H}}}

\def
\een{\end{displaymath}}

\def\R{\mathbb R}
\def\A{\mathbb A}
\def\W{\mathbb W}

\makeatother

\def\cH{{\mathcal{H}}}

\def\R{\mathbb R}

\def\be{\begin{equation}}\def\ee{\end{equation}}\def\f{\frac}
\def\A{\mathbb A}\def\bea{\begin{eqnarray}}\def\eea{\end{eqnarray}}
\def\ben{\begin{displaymath}}
\def\ba{\begin{array}{c}}\def\bal{\begin{array}{l}}\def\ea{\end{array}}
\def\dn{\mathrm{dn}}
\def\R{\mathbb R}\def\cH{{\mathcal{H}}}

\def\een{\end{displaymath}}

\begin{document}
	
	\title{
		{\bf Supersymmetric construction of self-consistent condensates in large $N$ GN model:  \\ solitons on finite-gap potentials }
	}
	
	\author{\textsf{Adri\'an Arancibia}
			\\	[4pt]
		{\small \textit{ Instituto de Matem\'atica y F\'isica, Universidad de Talca, Casilla 747, Talca, Chile}}\\
		\sl{\small{E-mail: adaran.phi@gmail.com} }}
	\date{}
	\date{}
	
	\maketitle	
	\begin{abstract}
	In the present work, the set of stationary solutions of the Gross-Neveu   model in  't Hooft limit is extended.
	Such extension  is obtained by striving a hidden supersymmetry associated to disconnected sets of stationary solutions. How the supersymmetry arises from the Darboux-Miura transformations between Lax pairs of the stationary  modified Korteweg-de Vries and the stationary Korteweg-de Vries hierarchies is shown, associating the correspondent superpotentials to self-consistent condensates for the Gross-Neveu model. 
	\end{abstract}
\begin{multicols}{2}
\section{Introduction}
The Gross-Neveu model (GN) \cite{GN}  corresponds to a quantum field theory for nonlinear interacting fermions without mass.  The model presents some interesting properties: dynamical mass generation, asymptotic freedom and spontaneous breaking of symmetry. 
Models with interacting fermions and self-consistent condensates have been used to describe a long variety of phenomena related to soliton physics, kinks and breathers. Especially in particle physics \cite{NJL, RaWil}, superconductivity \cite{FUFE,Peierls,CasNar} and conducting polymer models \cite{HKScSu,CaBi,SaxBish}, among other areas \cite{BDT,GBGD}. 

In the last 40 years, non-perturbative solution methods for this field theory in the   't Hooft limit (or large $N$ limit) have been studied in detail.
The first analytical solutions in this direction were obtained by applying the inverse scattering method \cite{DHN,JFShHi}, which allowed to relate fermionic condensates to superpotentials of  pairs of reflectionless systems of the Schr\"odinger type in 1 + 1D, thus bringing to light
 a  hidden nonlinear $N=4$ supersymmetry in the statio\-nary sector of the GN model \cite{GDJF,FCGDMP,PAM}.   
The inverse sca\-tte\-ring method has allowed the generation of an infinite fa\-mi\-ly of semi-classical condensates, corresponding to one-gap (massive) Dirac  potentials with solitary defects in its spectrum \cite{JF}.

A more general method to construct analytical solutions was found by applying the series expansion of Gorkov resolvent \cite{GBGD}, obtaining as a more general condition that the GN model semi-classical stationary condensates must be solutions of the inhomogeneous stationary modified  Korteweg-de Vries hierarchy (s-mKdVh). This led to the construction of exact  periodic inhomogeneous  condensate solutions known as kink crystals and kink-antikink crystals, which have already been found as self-consistent condensates through the Hartree-Fock  approximation \cite{ThiesLame,DT1,DT2}. Such condensates correspond to  two- and three- gap potentials for the Dirac Hamiltonian or Bogoliubov-de Gennes operator in 1 + 1D.
 
 The most important results in this paper are: 
 \begin{itemize}
 	\item  The extension of the stationary scalar condensates for the Gross-Neveu model to the most general form, as finite-gap Dirac potentials with solitonic defects. Such extension is achieved by using the supersymmetry hidden in the system. The supersymmetric method presented below allows to evade the inverse scattering approach and to algebraically construct  infinite families of extended Schr\"odinger Hamiltonians with central charge and  nonlinear $N=4$ supersymmetry starting from the exactly solvable finite-gap Schr\"odinger systems.  When one pair of fermionic integrals are of order one, it is possible to identify any of such fermionic integrals and the  central element of the respective superalgebra with the Lax pair formulation of an equation in the  s-mKdVh. Thus, the stationary condensate solutions of the GN model are identified with the superpotentials that define the set of fermionic integrals of order one.
 	The nonlinear $N=4$ hidden supersymmetry is observed by using the Darboux-Miura transformations and the Lax pair formalism of the s-KdVh and the s-mKdVh.
 	
 	\item From the recurrence relations in the construction of the s-mKdVh and  the  algebro-geometric formalism, the self-consistency equations  that fix the occupation of the allowed states by the fermions of different flavors for each condensate are  written in form of a consistent system of equations  with more unknowns than equations, therefore there are an infinite number of solutions. As an example, a special case of ground state that allows the decoupling of the occupations among bound states is studied. For this example the occupation of each bound state  only depends on its energy and the occupation of the states in the spectrum of the finite-gap background.
 	
 \end{itemize}

This work is structured as follows: in Sec. \ref{S1}, by means of the method of series expansion of the Gorkov resolvent, the self-consistent condensates of the GN model are identified with solutions of the  s-mKdVh. 
In Sec. \ref{S2}, the s-KdVh  and one useful  Lax pair formulation of it are summarized, showing their relation to symmetries in Quantum Mechanics. How to obtain s-mKdVh solutions starting from s-KdVh solutions will also be shown.
In Sec. \ref{S3}, the contents of the previous section are connected to Crum-Darboux spectral transformations   and  their relation to supersymmetric Quantum Mechanics is introduced.  
In Sec. \ref{S4}, algebro-geometric/finite-gap solutions of the  s-KdVh are presented  in Its-Matveev form. Through Crum-Darboux transformations, infinite families of finite-gap potential with soliton defects are obtained. Two types of spectral transformations are stu\-died from finite-gap potentials, which in Sec. \ref{S5}, will be related to three types of solutions of the GN model with different spectral characteristics. Finally, in Sec. \ref{S6},  the self-consistency  conditions of the stationary condensates of the GN model will be studied. As an example, the condensate with solitonic potential well defects on a two-gap background will be detailed. 

\section{The GN model, series expansion of the Gorkov resolvent and the s-mKdVh}\label{S1}
The GN model is described by the Lagrangian 
\begin{equation}\label{gn}
	{\cal L}_{\rm GN} = \sum_{j=1}^N\bar{\psi}_j i\partial \!\!\!/ \psi_j + \frac{g^2}{2} \left( \sum_{j=1}^N\bar{\psi}_j \psi_j \right)^2 
	\,,
\end{equation} 
where $\psi_j$, for $j=1,2,\ldots, N$, correspond to $N$ fermions of different flavors.
A bosonization is allowed for this model, where the bosonic field corresponds to the fermionic condensate $\Delta=-g^2 \left( \sum_{j=1}^N\bar{\psi}_j \psi_j\right)$. 

Through the path integral approach, an effective action $S_{\rm eff}$ for $\Delta$ can be obtained, 
\begin{equation}\label{effective}
S_{\rm eff}=-\int \frac{\Delta^2}{2 g^2} d^2x -i\,N\,\ln\det\left[i\, \partial \hskip -6pt / \,
-\Delta \right]\,.
\end{equation}
At the   't Hooft limit, where $N\rightarrow \infty$ and $g^2N \sim 1$, it is possible to use the saddle point method to ensure the convergence of the  two-point propagator associated with (\ref{gn}). The convergence happens for the minimums of (\ref{effective}). The variation of such action for the stationary case yields the following consistency equations
\begin{align}\label{consis}
	\Delta(x)=-iNg^2{\rm tr}_{D,\cE}\left[\gamma^0R(x; \cE)\right]\, ,
\end{align}
where $R(x; \cE)\equiv \langle x| (H^D-\cE)^{-1}|x\rangle$ is a $2\times 2$ matrix known as the Gorkov diagonal resolvent of the Bogoliubov-de Gennes operator or Dirac Hamiltonian in 1+1D,
\begin{align}\label{bdg1}
H^D	=
	\left(
	\begin{array}{cc}
		-i\frac{d}{dx}&\Delta(x)  \\
		\Delta(x) & i\frac{d}{dx}
	\end{array}
	\right), \quad H^D \Psi=\cE\Psi \, ,
\end{align}
$\gamma_0=\sigma_1$. Thus, the solutions of (\ref{consis})
correspond to the semi-classical solutions of the GN model.
At this point, it is possible  to connect with the Hartree-Fock treatment of the GN model, where the consistence equation for the condensate $\Delta$ is related  to the population of fermion flavors in the form 
\be
\Delta(x)=-g^2 \left( \sum_{occ}  \bar{\psi} \psi\right),
\ee
where $\sum_{occ}$ corresponds to the sum over all occupied states in the spectra of (\ref{bdg1}), taking into account the flavor degeneration.

By exploiting the properties of the resolvent in (\ref{consis}), a general approach  to construct analytical solutions for the model was found\cite{GBGD,BDT}. The resolvent $R(x; \cE)$ satisfies the following algebraic properties  $R=R^\dagger$, ${\rm tr}_D \left(R\sigma_3\right)=0$, ${\rm det}\, R=-\frac{1}{4}$ and also satisfies the Dickey-Eilenberger equation \cite{resol} 
\begin{eqnarray}\label{DicEil}
\frac{\partial}{\partial x}R \, \sigma_3
&=&i\, \left[
\begin{pmatrix}
\cE&-\Delta \cr
\Delta & -\cE
\end{pmatrix}, R \,\sigma_3
\right].
\end{eqnarray}
The power series expansion of the Gorkov resolvent on  ener\-gy variable,  $R=\sum_n^\infty r_n(x)/\cE^n$, can be truncated in order to find analytic solutions for the condensate $ \Delta(x) $. In this case the resolvent  takes the form
\begin{eqnarray}\label{rn}
R_n(x; \cE)={\mathcal N}(\cE)\sum_{l=0}^n \cE^{n-l}\begin{pmatrix}
\hat{g}_l(x) & \hat{f}_{l-1}(x) \cr
\hat{f}_{l-1}^*(x) & \hat{g}_l(x)
\end{pmatrix}
\label{rexpmatrix},
\label{resolvent}
\end{eqnarray}
under the condition 
\begin{equation}\label{cond}
\hat f_n=0,
\end{equation}
the latter is known as the $n$-th equation of the s-mKdVh,
where $\hat{g}_l(x)$ and $\hat{f}_l(x)$ are completely defined  by  (\ref{DicEil})
in the following recursive form
\bea
\hat{f}_l&=&-\frac{i}{2}\hat{f}_{l-1}^{\,\prime}+\Delta \,
\hat{g}_l,\nonumber\\ \hat{g}_l &=& i\int\left(\hat{f}_{l-1}-\Delta\, \hat{f}_{l-1}^*
\right) dx +c^D_l,\nonumber
\eea
\begin{equation}
\hat{f}_{-1}=0,\quad\hat{f}_{0}=\Delta(x),\quad\hat{g}_0=c^D_0=1,
\end{equation}
where $c^D_l$ are real integration constants, $c^D_{2j+1}=0$, $j,l\in\mathbb{N}_0$. 

The truncation condition (\ref{cond}) defines $\Delta(x)$ as a solution of the s-mKdVh. The first five equations in the hierarchy correspond to
\bea\label{hfl}
\hat{f}_{-1}(x)&=&0,\nonumber\\
\hat{f}_{0}(x)&=&\Delta(x),\nonumber\\
\hat{f}_{1}(x)&=&-\frac{i}{2}\Delta',\nonumber\\
\hat{f}_{2}(x)&=&-\frac{1}{4}(\Delta''-2\Delta^3)
+c^D_2\Delta,\nonumber\\
\hat{f}_{3}(x)&=&\frac{i}{8}(\Delta'''-6\Delta^2\Delta')
-\frac{i c^D_2}{2}\Delta'.\nonumber
\eea
This hierarchy of equations corresponds to integrable systems and can be solved by algebro-geometric methods. This is because the equations in the s-mKdVh  allow Lax pair formulation, which corresponds to write (\ref{cond}) as two commutating operators $[H^D,P^D]=0$, where $P^D$ is a $2\times 2$ matrix differential operator of order $n$ and takes the role of the Lax-Novikov integral of the $H^D$ Dirac Hamiltonian.
For the s-mKdVh, the Lax integral $P^D$ takes the form 
\begin{equation}\label{mLN}
P^D_{n}=  \sum_{\ell=0}^{n}\begin{pmatrix}
\hat{g}_\ell(x) & \hat{f}_{\ell-1}(x) \cr
\hat{f}_{\ell-1}^*(x) & \hat{g}_\ell(x) 
\end{pmatrix} \sigma_3  H^{D}{}^{n-\ell },
\end{equation}
for which the Lax equation in the stationary case  is 
\begin{equation}
[P^D_{n},H^D]=  \begin{pmatrix}
0 & 2\hat{f}_{n}(x) \cr
-2\hat{f}_{n}^*(x) &0
\end{pmatrix}={\bf 0} ,
\end{equation}
This formulation relates stationary solutions $\Delta(x)$ of the mKdVh  to  scalar potentials for one dimensional Dirac  Hamiltonian operators that have  Lax-Novikov integral of motion.
In addition, the interpretation of those Dirac Hamiltonians as Bogoliubov-de Gennes  (\ref{bdg1}) operators relates the $\Delta(x)$
potentials  to the stationary solutions of the GN model.

The coefficients $c^D_k$ are related to the edges of the spectrum of Hamiltonian operator  $H^D$,
\bea\label{sHD}
\sigma(H^D)&=&(-\infty,\cE_0]\cup[\cE_1,\cE_2]\cup \ldots\nonumber\\
&&\cup[\cE_{2j-1},\cE_{2j}]\cup [\cE_{2n+1},\infty),
\nonumber
\eea
$\cE_{2j-1}\leq \cE_{2j}$, in the form
\begin{equation}
c^D_k=\sum_{\substack{i=j_0,j_1,...,j_n=0 \\ j_0+j_1+..+j_n=k}}^{k} 2^{-2k}
\prod_{i=0}^{2n+1} \frac{(2 j_i )!}{(j_i !)^2(2 j_i-1)} (\cE_i){}^{j_i}.
\end{equation}

The self-consistency equation (\ref{consis}) corresponds to a system of equations that defines the occupation of  each physical state of the spectra of  $H^D$  by the different flavors.

Another important behavior of the Lax pair operators is a Burchnall-Chaundy type relationship between matrix differential operators, which relates powers of the Lax pair operators in the following form
\begin{equation}
P^{D}_{n}{}^{2}=\cP_{n,BC}(H^D)=\prod_{\ell=0}^{2n-1}(  H^D-\cE_\ell ),
\end{equation}
which defines the eigenvalues $z^D$ of $H^D$ and $y^D$ of $P^D$ over a hyper-elliptic curve 
\begin{equation}
(y^D){}^2=  \prod_{\ell=0}^{2n-1}(  z^D-\cE_\ell ).
\end{equation}
The latter relation is in the basis of  the algebro-geometric solution method for the s-mKdVh.

The normalization constant ${\mathcal N}(\cE)$ of the dia\-go\-nal resolvent (\ref{rn}) is  defined by the condition ${\rm det}R=-\frac{1}{4}$ as follows 
\be\label{detN}
{\rm Det}(R_n(x;\cE))={\mathcal N}(\cE)^2\cP_{n,BC}(\cE)=-\frac{1}{4}.
\ee


\section{The s-KdVh, symmetries in Quantum Mechanics and s-KdVh $\rightleftharpoons$ s-mKdVh Miura transformation }\label{S2} 
The supersymmetric method of construction of new condensates for the GN model is based on the Miura transformation between solutions of the mKdVh and the KdVh. In order to introduce this transformation, first it is necessary to summarize the s-KdVh, and then how to obtain solutions of the s-mKdVh from pairs of solutions of the s-KdVh will be shown.

The s-KdVh corresponds to a set of nonlinear integrable systems,  and its equations are recursively defined as follows
\be
f_{\ell,x}=-\frac{1}{4}f_{\ell-1,xxx}+uf_{\ell-1,x}+\frac{1}{2}u_{x}f_{\ell-1},
\ee
$f_0=1,$ being the equations of the s-KdVh
\be
2f_{\ell,x}=0,
\ee 
Explicitly, one finds
\bea\label{fl}
2f_{0,x}&=&0,\nonumber\\
2f_{1,x}&=&u_x\nonumber\\
2f_{2,x}&=&-\frac{1}{4}\left(u_{xxx}-6uu_x-4c_1u_x\right)\nonumber\\
2f_{3,x}&=&\frac{1}{8}\left(16u_{xxxxx}-5u_xu_{xx}-5u_x^2-5uu_{xx}\right.\nonumber\\
& &+15u^2u_x-2c_1\left(u_{xxx}-6uu_x\right)\nonumber\\
& &\left.+8c_2u_x\right),\nonumber\\
\vdots& & \nonumber
\eea
where $c_\ell$ are real valued integration constants.

These equations allow Lax pair formulation in the form
\be\label{laxeqKdV}
2f_{\ell+1,x}=i[P_{2\ell+1},H],
\ee
where $[.,.]$ is the usual commutator, and
\be\label{HSch}
H=H(u)=-\frac{d^2}{dx^2}+u,
\ee
corresponds to the stationary Schr\"odinger operator in one dimension, $\hbar=2m=1$, and the operator
\bea\label{P}
P_{2\ell+1}&=&P_{2\ell+1}(u,\partial\sigma(H(u)))\\
&=&-i\sum^{\ell}_{j=1}\left(f_{\ell-j}\frac{d}{dx}-\half f_{\ell-j,x}\right)H^\ell,\nonumber
\eea
is called the Lax operator and $\partial\sigma(H(u))$ is the border of the  physical spectrum of $H(u)$ that defines the coefficients $c_j$ in  $P_{2\ell+1}$.


The Lax equation  $[H,P_{2\ell+1}]=0$ describes an odd order integral of motion for each Schr\"odinger  Hamiltonian associated to solutions $u$ of the s-KdVh. The order of this integral, called the Lax-Novikov integral, depends on the number of allowed bands and bound states in the spectrum of such Hamiltonian. 

 The constants $c_j$ in (\ref{fl}) are defined in function of the energies 
$E_m\in\partial\sigma(H(u))$, 
\begin{align}
	c_k=-\sum_{\substack{i=j_0,j_1,...,j_{2g}=0 \\ j_0+j_1+..+j_{2g}=k}}^{k} 2^{-2k}
	\prod_{i=0}^{2g} \frac{(2 j_i )!}{(j_i !)^2(2 j_i-1)} E_i^{j_i}
\end{align}
where $k=1,..., \ell$ and $c_0=1$. 

There exists a transformation between the  mKdVh solutions and the KdVh solutions, this is called Miura transformation. 
The Miura transformation is defined by  
\be\label{udefv}
u=v^2-v_x,
\ee
if $v$ is any s-mKdVh solution, there is a $\hat{f}_{2n-1}$ such that
\begin{equation}\label{mKdV}
\hat{f}_{2n-1}(v)=0,
\end{equation} 
then due to the identity 
\begin{equation}\label{d-mKdV}
f_{n,x}(u)=i(2v-\partial_x)\hat f_{2n-1}(v),
\end{equation} 
$u$ is a s-KdVh solution, 
\begin{equation}\label{KdV}
f_{n,x}(u)=0.
\end{equation}
Note that the inverse affirmation is not correct.

The s-mKdVh  is invariant under the change $v\rightarrow -v$, hence the transformation (\ref{udefv})
allows to define 
\begin{equation}\label{udefv+}
u^+=v^2+v_x,
\end{equation}
and
\begin{equation}\label{udefv-}
u^-=v^2-v_x,
\end{equation}
where $u^{\pm}$ are both  s-KdVh solutions dependent on $v$. 
From another perspective:  let two functions
$u^+(x)$ and $u^-(x)$  that satisfy the same equation in s-KdVh and depend on a function $v(x)$ as in (\ref{udefv+}) and
(\ref{udefv-}) respectively. In this case,  $v(x)$ must satisfy \emph{simultaneously} (\ref{d-mKdV}) for $v$ and for $-v$. 
By adding these two equations the following identity is obtained $4v\hat f_i=0$, which implies that $v$ must satisfy s-mKdV equation (\ref{mKdV}). 
This frame is in the basis of the hidden supersymmetry of the stationary sector of the GN model. It is natural to ask: why change the problem from the search of one solution of the mKdVh to the search  of two connected solutions of the KdVh?. In the next sections, it is shown how by starting from an initial Schr\"odinger potential $u^+ $ and its eigenstates, it is possible to construct through Darboux transformation a supersymmetric pair $u^-$ and a superpotential $v(x)$. A characteristic of the Darboux transformation is to keep the symmetries, so if $u^+$ has a Lax-Novikov integral (i.e. a s-KdVh solution) then $u^-$ also has its respective Lax-Novikov integral (i.e. another s-KdVh solution). By observing intertwining operators that generate such transformations ($A=\frac{d}{dx}+v(x)$), families of superpotentials $v(x)$ will be found, which will be solutions of the s-mKdVh and candidates to self-consistent stationary condensates of the Gross-Neveu model.   


\section{Crum-Darboux transformations and supersymmetry in 1+1D }\label{S3}

The system of equations (\ref{udefv+}) and (\ref{udefv-}) hides a supersymmetry in its structure. From the point of view of the Lax pair formulation, $u^+$ defines the Schr\"odinger Hamiltonian $H^+=-\frac{d^2}{dx^2}+u^+$  and  $u^-$ defines the Schr\"odinger Hamiltonian $H^-=-\frac{d^2}{dx^2}+u^-$,  while $v$ defines the Dirac operator $H^D=-i\frac{d}{dx}\sigma_3+v\sigma_1$. To simplify, an unitary transformation to $H^D$ is done, which defines the operator $\mathcal{Q}_{1}=e^{-i\frac{\pi}{4}\sigma_1}H^De^{i\frac{\pi}{4}\sigma_1}$, such that
 \begin{equation}\label{Q}
 \mathcal{Q}_{1}=\begin{pmatrix}0&A\\
 A^\dag
 &0\end{pmatrix}=\begin{pmatrix}0&\frac{d}{dx}+v\\
 -\frac{d}{dx}+v
 &0\end{pmatrix},
 \end{equation}
this operator $Q_1$ plays the role of the square root of 
\bea
\cH-E_\star&=&\begin{pmatrix}H^+&0\\
	0
	&H^-\end{pmatrix}\\&=&-\frac{d^2}{dx^2}+v^2+\sigma_3 v_x\nonumber\\
&=&\mathcal{Q}_1^2=\half\{\mathcal{Q}_1,\mathcal{Q}_1\}\nonumber.
\eea
The operator $\cH$  can be interpreted as an extended Schr\"odinger operator and corresponds to the Witten Hamiltonian of supersymmetric Quantum Mechanics in 1+1D. 
The  Hamiltonian $\cH$  presents two fermionic integrals in the form  $\cQ_1$ and $\cQ_2=i\sigma_3\cQ_1$
for the grading operator $\Gamma=\sigma_3$. The Lie superalgebra of these integrals of motion takes the following form 
\be
[\mathcal{H} ,\mathcal{Q}_{a}]=0,\quad
\{\mathcal{Q}_{a},\mathcal{Q}_{b}\}=2\delta_{ab}(\mathcal{H}-E_\star),
\ee
where $a,b=1,2$.

Besides, it is possible to observe that $\Psi^\pm(x)=\exp{\left(\pm\int^x_{x_0}{\rm d}x'v(x')\right)}$ are eigenstates (not necessarily physicals) of $H_\pm$ with energy $E_\star$, or equivalently
\be
H^\pm\Psi^\pm(x)=E_\star\Psi^\pm(x).
\ee
Starting from the state $\Psi^\pm(x)$ of $H^\pm$, $u^\mp(x)$ is defined as a Darboux transformation of $u^\pm(x)$ in the form
\be
u^\pm(x)\rightarrow u^\mp(x)=u^\pm(x)-2\left(\ln(\Psi^\pm(x))\right)'', 
\ee
this type of transformations plays an important role in the theory of integrable systems, spectral analysis and soliton systems. In the context of such transformations, the components of the fermionic integrals $\mathcal{Q} $, defined in (\ref{Q}),  $A= \frac{d}{dx}+ v(x)$ and $A^\dag=-\frac{d}{dx}+ v(x)$ (Hermitian conjugated of $A$), are known as intertwining operators between $H^+$ and $H^-$. Due to the factorizations $H^+-E_\star=A  A^\dag$ and $H^--E_\star=A^\dag A$ 
the following intertwining relations are fulfilled
\be\label{intrel}
AH^-=H^+A,\quad A^\dag H^+=H^-A^\dag.
\ee

These identities play a fundamental role in the solution of spectral problems of high complexity, since these allow to obtain the spectrum of  $H^\mp$ from the spectrum of  $H^\pm$, and vice versa. If $\Psi^\pm(x,E)$ is a state of $H^\pm$ with energy $E$, then the intertwining  relations (\ref{intrel}) imply that  $A^\dag\Psi^+(x,E)$ is a state of $H^-$ with energy $E$, while $A$ performs the reverse mapping, $A\Psi^-(x,E)$ is a state of $H^+$ with energy $E$. There will be some problems with the mapping of the states $\Psi^\pm(x)$, since $A$ can be written as
\be   
A=\Psi^-(x)
\frac{d}{dx}\frac{1}{\Psi^-(x)},  
\ee
so it annihilates  $\Psi^-(x)$ among the eigenstates of  $H^-$, i.e. $A\Psi^-(x)=0$,
while $A^\dag$ annihilates the state $\Psi^+(x)$, of same energy, among the eigenstate spectrum of $H^+$. In fact by definition  
\be
A^\dag=-\Psi^+(x)\frac{d}{dx}\frac{1}{\Psi^+(x)},
\ee
$\Psi^+(x)=\frac{1}{\Psi^-(x)}$,  $A^\dag\Psi^+(x)=0$. At this point, it is interesting to note that if $\Psi^\pm(x)$ is a  concave (convex) state without zeros for $x\in\R$ then $\Psi^\mp(x)$ is a bounded eigenstate of $H^\mp$. In this case $u^\mp(x)$ shows a solitonic defect in form of potential well, which supports such bound state. 

Darboux transformation is generalized by Crum-Darboux transformation \cite{Crum,MatSal}. Such transformation  corresponds to the application of successive Darboux transformations and  induces a formulation of non-linear supersymmetry in quantum mechanics.
An  order $n$ Crum-Darboux transformation to Schr\"odinger operator 
$H_0=-\frac{d^2}{dx^2}+V_0(x)$, results in a new operator 
\bea\label{Hn}
H_n&=&-\frac{d^2}{dx^2}+V_n(x),\\
 V_n&=&V_0-2\frac{d^2}{dx^2}\log \W_n\, ,\nonumber
\eea
where  $\W_n$ is the Wronskian of  $n$ formal eigenstates $\psi_j$  of  $H_0$,
$H_0\psi_j=E_j\psi_j$, $E_i\neq E_j$, 
\bea
\W_n&=& \W(\psi_1,\ldots,\psi_n)=\det \mathcal{A},\\
\mathcal{A}_{ij}&=&\frac{d^{i-1}}{dx^{i-1}}\psi_j,
\quad i,j=1,\ldots, n\,\nonumber
\eea
The eigenstates  $\Psi_0(x;E)\neq \psi_j$  of $H_0$,
$H_0 \Psi_0(x;E)=E\Psi_0(x;E)$, are mapped to eigenstates 
$\Psi_n(x;E)$ of  $H_n$, $H_n \Psi_n(x;E)=E\Psi_n(x;E)$,
through the fraction of Wronskians,
\be
\Psi_n(x;E)=\frac{\W(\psi_1,\ldots,\psi_{n},\Psi_0(E))}{\W_{n}},  
\ee
where $\W_0=1$ has been chosen. It is possible to introduce the first-order differential intertwining operators
\bea\label{An}
	A_n&=&
	\f{d}{dx}+\mathcal{W}_n\,,\\
	\mathcal{W}_n&=&
	-\frac{d}{dx}\log \W_n+\frac{d}{dx}\log \W_{n-1}\, .\nonumber
\eea
These operators and their conjugate factorize $H_{n-1}$  and $H_n$ in the form  
$A_n^\dag A_n=
H_{n-1}-E_{n},$
$A_nA_n^\dag=
H_n-E_{n},   
$
and intertwine them as follows
$A_nH_{n-1}=H_nA_n\,,$ $A_n^\dagger H_n=H_{n-1}A_n^\dagger\,.$

The intertwining operator  $A_n$  can equivalently be represented in the form
\bea\label{Ajdef}
	A_j&=&(A_{j-1}\ldots A_1\psi_j)\frac{d}{dx} \frac{1}{(A_{j-1}\ldots
		A_1\psi_j)}\nonumber\\&=&\frac{d}{dx} -\left(\frac{d}{dx} \ln(A_{j-1}\ldots
	A_1\psi_j)\right)\,,
\eea
where $A_1=\psi_1 \frac{d}{dx} \frac{1}{\psi_1}$ and $A_{j-1}\ldots A_1\psi_j$ is an eigenstate of eigenvalue  $E_j$ for $H_{j-1}$, any other formal eigenstate $\Psi_{j-1}(E)$ of $H_{j-1}$, 
$H_{j-1}\Psi_{k-1}(E)=E\Psi_{n-1}(E)$, is mapped for $A_j$ in the eigenstate  $\Psi_j(E)=A_j\Psi_{j-1}(E)\,$ of  $H_j$ with same eigenstate as $H_{j}\Psi_{j}(E)=E\Psi_{j}(E)$.

In this way, it is possible to intertwine $H_n$ and  $H_0$ using the order $n$ intertwining operator $\A_n\equiv A_n\ldots
	A_1\,,$ $\A_nH_0=
	H_n\A_n$, $\A_n^\dagger
	H_n=H_0\A_n^\dagger$.
If $\W_n\neq0$ and $V_0(x)$ is nonsingular, for $x\in  \R$, 	then the extended system $\cH={\rm diag}(H_0,H_n)$ is characterized by a nonlinear supersymmetry dependent on the scattering  data of the eigenstates used in Crum-Darboux transformation. In the superalgebra there exist two nilpotent $Z_2$-odd anti-diagonal supercharges  $Q_+=\A_n^\dag\sigma_+=\frac{1}{2}(Q_2+iQ_1)$ and
$Q_-=\A_n\sigma_-=Q_+^\dagger$,  $[Q_\pm, \cH]=0$, $Q_\pm^2=0$, where
$\sigma_\pm=\frac{1}{2}(\sigma_1\pm i\sigma_2)$. These generate a nonlinear Lie superalgebra in the form 
$\{Q_a,Q_b\}=2\delta_{ab}\prod^n_{\ell=1}(\mathcal{H}-E_\ell)$.

This supersymmetric representation shows a spontaneous breaking of symmetry, which depends on the spectral data of the chosen  $\psi_i$ states. The $H(u)$ Schr\"odinger operator has an order two  formal degeneration for each energy level, so the election of the states $\psi_i$ is arbitrary, and in general, it is a linear combination between the pair of linearly independent states of same energy. There are states that through Darboux transformations produce i) nonlinear phase shift in the initial potential, ii) one defect in initial potential with form of soliton potential well and/or iii) singu\-la\-ri\-ties.  In addition to the above, special elections of pairs of states that produce singularities 
can altogether generate one or two soliton defects,  supporting bounded states in the forbidden gaps of the spectrum of the initial potential. 

It is possible to differentiate between three supersymmetric frames: exact, broken and partially broken supersymmetries. 
Let $H_n$ and $\psi_i$ be defined as in \ref{Hn}, the supersymmetry associated to $\cH$ is exact if $H_n$ ($H_0$) has a normalizable ground state of energy $E_i$ lower than the energy of the ground state of $H_0$ ($H_n$). In this case, $\cH$ supports a ground state in the form $\Psi_0=(0,\A_n\psi^\star_i)^T$ ( $\Psi_0=(\psi_i,0)^T$), where $\psi^\star_i$ is a state of $H_0$ of energy $E_i$ linearly independent to $\psi_i$, thus the ground state $\Psi_0$ is annihilated for all generators of supersymmetry $Q_j\Psi_0=0$, $j=1,2$. 

The supersymmetry is broken if  $H_0$ and $H_n$ have both normalizable ground states of same energy $E_i$, such that $\cH$ has two normalizable ground states 
$\Phi_{0,0}=(\sqrt{\prod^n_{\ell=1}(E_0-E_\ell)}\psi_0,0)^T$ and $\Phi_{0,n}=(0,\A_n\psi_0)^T$, where $\psi_0$ is the bound state of $H_0$. In this case the generators of supersymmetry do not annihilate the ground states, rather they transform them one into the other $Q_-\Psi_{0,0}=\sqrt{\prod^n_{\ell=1}(E_0-E_\ell)}\Psi_{0,n}$ and $Q_+\Psi_{0,n}=\sqrt{\prod^n_{\ell=1}(E_0-E_\ell)}\Psi_{0,0}$,  $Q_-\Psi_{0,n}=Q_+\Psi_{0,0}=0$. 

A third case arises if the lower energy state in Crum-Darboux transformation $\psi_i$ corresponds to the lower energy edge of bands, in this case $\psi_0=\psi_i$ and the supersymmetry generators annihilate both states $Q_\pm\Phi_{0,a}=0$, $a=0,n$.
About the latter case, it is necessary to say that the finite-gap structure of $V_0$ expands the number of supersymmetry generators from $N=2$ to $N=4$. There is a central extension of the superalgebra due to the Lax-Novikov integral present in finite-gap systems. This is called a partially broken supersymmetry because the initial two fermionic integrals annihilate the two bound states while the additional  two do not \cite{5A}. 

The central extension is also possible in the first two cases but it is essential in the definition of the third one. 
 
An important behavior occurs when Crum-Darboux transformation takes place, for it makes possible to preserve the symmetries of the initial system. For example: let $H$ be a Hamiltonian with integral of motion $P$, $[H,P]=0$, a Crum-Darboux transformation to $H$ allows to define a Hamiltonian $H'$ and an intertwining  operator  $A$ between $H$ and $H'$, $AH=H'A$ and $A^\dag H'=HA^\dag$, then through Darboux dressing to $P$ it results that $P'=APA^\dag$ is an integral of $H'$. In simple steps it is proved that $[H',P']=0$, which ensures that by means of Crum-Darboux transformation a solution of the s-KdVh ($[H,P]=0$) yields another solution of the s-KdVh ($[H',P']=0$). The order of the equation in the hierarchy solved by the transformed system depends on the order of $P'$ and it can be equal,  lower or higher than the initial solution order. This is possible due to an order reduction mechanism that relies on the spectral data of $H$ and the states used in Crum-Darboux transformation. In this direction, sometimes it is possible to reduce the operator $P'$ into a lower order operator $ \tilde{P}'$ due to an identity in the form $P'=\prod_{i}(H-\lambda_i) \tilde{P}'$ which ensures that if $P'$ is an integral, then $ \tilde{P}'$ also is. On the other hand, if $u$ is a solution of one equation of the s-KdVh then auto\-ma\-ti\-ca\-lly  it is solution of infinite equations of higher order of the hierarchy. In other words, if $u$ is a solution of any equation of the hie\-rar\-chy that defines a Lax pair $H$ and $P$, $[H,P]=0$, then it is always possible to construct a new Lax pair  $H$ and $\tilde P=P\prod_{j=1}^n (H-\lambda_j)$, $[H,\tilde P]=0$, where $\lambda_j\in \R$ and $n$ a positive integer, that corresponds to the Lax pair formulation of a higher order equation in s-KdVh.

\section{Algebro-geometric solutions of the  s-KdVh and their Crum-Darboux transformations}\label{S4} 

In this section the Its-Matveev formula of the algebro-geometric finite-gap solutions of the s-KdVh is presented, moreover, two types of spectral transformations are studied and infinite families of finite-gap potential with solitonic defects are obtained. 

The Lax pair of KdVh  $H_{g,0}=H(u_{g,0})$ and $P_{2g+1}=P_{2g+1}(u_{g,0},\partial\sigma(H_{g,0}))$ satisfy the Burchnall-Chaundy relationship \cite{BurCha}
\begin{equation}\label{BC}
P^2_{2g+1}=\prod_{i=0}^{2g}(H-E_i)\,,
\end{equation}
which relates the eigenvalues $y$ of $P_{2g+1}$ to the eigenvalues  $z$ of $H_{g,0}$ through the hyper-elliptic curve 
\begin{equation}\label{HEC}
y^2=\prod_{i=0}^{2g}(z-E_i)\,.
\end{equation}

The algebro-geometrical method allows to find  the finite-gap solutions in the form of second derivatives of logarithms of Riemann theta functions.
The Its-Matveev formula \cite{ItsMat,GesHol} for potentials with $g$-gaps, is given by
\be\label{itsmat}
u_{g,0}(x)=-2\frac{d^2}{dx^2}\ln(\theta(x{\bf v}+{\boldsymbol \phi},\tau))+\Lambda_0, 
\ee
with ${\bf v},{\boldsymbol \phi}\in 
\mathbb{C}^g$. 

  The eigenstates for the Hamiltonian associated  to (\ref{itsmat})  are given in the form
\be\label{itsmat2}
\psi(r,x)=\frac{\theta(x{\bf v}+{\boldsymbol \phi}+{\boldsymbol \alpha}(r),\tau)}{\theta(x{\bf v}+{\boldsymbol \phi},\tau)}\exp
\left(-i x \xi(r)\right),
\ee
where $\theta$ is the Riemann Theta function of genus $g$ which presents a periodicity in the form  $\theta({\bf z}+{\bf a},\tau)=\theta({\bf z},\tau),
$ for ${\bf a}\in \mathbb{Z}^g$,
\be
\theta({\bf z},\tau)=\sum_{{\bf n}\in \mathbb{Z}^g}\exp\left(2\pi i <{\bf n},{\bf z}> +\pi i <{\bf n},{\bf n}\tau>\right),
\ee
with  ${\bf z}\in \mathbb{C}^g$ and  $\tau$  the mo\-du\-lar matrix.
The genus $g$ of the Riemann theta function corresponds to the number of  
band gaps in the spectrum of the associated Schr\"odinger operator \cite{Belo,GesHol}. 

The parameters in (\ref{itsmat}) and (\ref{itsmat2}) are completely defined by the curve (\ref{HEC}). The modular matrix is a $g\times g$ symmetric matrix with positive defined imaginary part, whose elements, as well as the components of ${\bf v}$ and the constant $\Lambda_0$, are uniquely determined by the energies of the edges of the spectrum of $H$, while ${\boldsymbol \alpha}(r)$ and $\xi(r)$ also depend on a point $r=(z,y)$ on the hyperelliptic curve (\ref{HEC}) such that $H\psi(r,x)=z\psi(r,x)$ and $P_{2g+1}\psi(r,x)=y\psi(r,x)$. On the other hand, ${\boldsymbol \phi}$ depends on the full spectral data of $H$ \cite{GesHol}. 

Crum-Darboux transformations to solutions in Its-Matveev form correspond to finite-gap systems with bound states in their forbidden bands.  
An equation in s-KdVh  $2f_{g+l+1,x}(u_{g,l}(x))=0$ with parameters  $c_\ell$, $\ell=0,\ldots,2g+2l$, defined by the energies $\partial\sigma(H_{g,l})=\{E_0,...,E_{2g}\}\cup(\cup_{i=1,\ldots,l}\{z(r_{i,1}),z(r_{i,1})\})$ 
has solutions with irreducible $P_{2g+2l+1}$ when $u_{g,l}$ takes the form
\end{multicols}
\rule{91mm}{0.1mm}
\be\label{ugl}
u_{g,l}(x)=u_{g,0}(x)-2\frac{d^2}{dx^2}\ln(\W(\psi_{a_{1,1},a_{1,2}}(r_{1,1},r_{1,2},x),\ldots,\psi_{a_{l,1},a_{l,2}}(r_{l,1},r_{l,2},x))),
\ee
\rule{95mm}{0mm}\rule{91mm}{0.1mm}
\begin{multicols}{2}
\noindent with
\bea\label{psiaa}
\psi_{a_{i,1},a_{i,2}}(r_{i,1},r_{i,2},x)=& &a_{i,1}\psi(r_{i,1},x)\\
&+&a_{i,2}\psi(r_{i,2},x)\nonumber
\eea
real functions, where $a_{i,1}$ and $a_{i,2}$ are $\mathbb{C}$ constants and $r_{i,1}$ and $r_{i,2}$ are elements in different charts of the  Riemann surface related to hyper-elliptic curve  (\ref{HEC}) with $z(r_{i,1})=z(r_{i,2})$ and $y(r_{i,1})=- y(r_{i,2})$, $z_{i}\neq z_j$, for $i\neq j$, $i,j=1,\ldots,l$. 

The s-KdVh solution $u_{g,l}(x)$ defines a Lax pair in the form
\bea
H_{g,l}&=&H(u_{g,l}(x)),\\
 P_{2g+2l+1}&=& P_{2g+2l+1}(u_{g,l}(x), \partial\sigma(H_{g,l})),\nonumber
\eea
where the Darboux dressing of the integral $P_{2g+1}$ yields the identity $P_{2g+2l+1}=\A_{l}P_{2g+1}\A_{l}^\dag,$
$\A_{l}=A_lA_{l-1}\cdots A_1$ with $A_j$ defined as in (\ref{An}) but changing $\psi_i\rightarrow \psi_{a_{i,1},a_{i,2}}(r_{i,1},r_{i,2},x)$.  

In order to perform a quantum mechanical interpretation, the  operator $ H $ must fulfill the role of Hamiltonian and $P_{2g+2l+1}$ the role of integral of motion. To ensure real valued observables it must be required that both $H$ and $P$ be  Hermitian operators without singularities in the real axis\footnote{It is not necessary to require this interval as an hermiticity condition (an example of this is the infinite potential well) but, given the nature of solitonic potentials, this is the most natural choice for present type of systems.}. 
It is necessary to demand that $u_{g,l}(x)$  have no singularities for $x\in\mathbb{R}$ and that its spectrum  be real, which implies $E_i,z_j,\Lambda_0\in\mathbb{R}$, for $i=0,\ldots,2g$ and $j=1,\ldots,l$. Under these conditions the operation $\dag$ corresponds to the Hermitian conjugation. 

To construct $u_{g,l}(x)$ non singular it is possible to choose  $u_{g,0}(x)$ and $u_{g,l}(x)-u_{g,0}(x)$  both non-singular or both singular but in the latter case the singularities of the first term must cancel the singularities of the second. To obtain  $u_{g,0}(x)$ singular, it is enough with a correct election of {\bf $\phi$}. Overall, this is possible when $z_i\in\sigma(H)^{c}$, where $c$ cor\-res\-ponds to the complement, i.e. $\psi_{a_i,a_2}(r_{i,1},r_{i,2},x)$ must be nonphysical states of $ H $. It is necessary to use the zeros theorem for the co\-rrect choice of sign of the ratio  $a_{i,1}/a_{i,2}$ for each energy $z(r_{i,1})$. Note that there are an infinite number of solutions. 

Soliton potential wells added by Crum-Darboux transformations to solution in Its-Matveev form  deform the shape of the initial finite-gap potential. It is pos\-si\-ble to obtain different shape types of  solitonic potential wells depending on which forbidden band supports the associated bounded states. The asymptotic behaviors of the Crum-Darboux transformation  correspond to phase shifts in the periodic structures of the initial potential. 


\subsection{ Types of Miura-Darboux transformations}
Two types of Darboux transformations to $u_{g,l}(x)$ are in\-te\-res\-ting. Their differences produce three types of solutions for the GN model. 

{\bf {Auto Darboux transformations}}: the real eigenstates $\A_{l}\psi(r,x)$ for $H_{g,l}$, $z(r)\leq E_i$, $E_i\in\partial\sigma(H_{g,l})$, correspond to nonsingular states of  modulated  exponential grow\-ing (decreasing), which can not generate bound states through Darboux transformation. As consequence of Its-Matveev formula,  Darboux transformation using $\A_{l}\psi(r,x)$ connects two shape invariant iso-spectral potentials, $u_{g,l}(x)$ and  $\tilde{u}^{r}_{g,l}(x)$. The potential $\tilde{u}^{r}_{g,l}(x)$ differs from $u_{g,l}(x)$ in phase shifts of the  positions of the solitons (due to changes in the coefficients $a_{l,i}$) and of the crystal structures that generate each band. The phase shift in crystal structures is given as follows  
\be\label{atds}
\delta{\boldsymbol \phi}=\tilde{{\boldsymbol \phi}}^{r}-{\boldsymbol \phi}={\boldsymbol \alpha}(r),
\ee
where  ${\boldsymbol \phi}$ is defined in (\ref{itsmat}) and ${\boldsymbol \alpha}(r)$ comes from the definition of $\psi(r,x)$ in (\ref{itsmat2}). ${\boldsymbol \phi}$ and $\tilde{{\boldsymbol \phi}}^{r}$ correspond to vectors whose components are the phases of the crystal structures in $u_{g,l}(x)$ and $\tilde{u}^{r}_{g,l}(x)$ respectively.

The operators
\bea\label{Xrl}
X_l(r)&=&\A_l\psi(r,x)\frac{d}{dx}\frac{1}{\A_l\psi(r,x)},\\ 
X_l(r)^\dag &=& -\frac{1}{\A_l\psi(r,x)}\frac{d}{dx}\A_l\psi(r,x),
\eea
intertwine 
$H_{g,l}=X_l(r)^\dag X_l(r)+z(r)$ and $\tilde{H}^{r}_{g,l}=X_l(r) X_l(r)^\dag+z(r)=-\frac{d^2}{dx^2}+\tilde{u}^{r}(x)$
in the form
$X_l(r)H_{g,l}=\tilde{H}^{r}_{g,l}X_l(r)$ and $X_l(r)^\dag\tilde{H}^{r}_{g,l}=H_{g,l}X_l(r)    ^\dag$, respectively. By being the initial and the transformed potentials iso-spectral by construction, these must be solutions of the same equation of the s-KdVh with exactly the same coefficients $c_\ell$. Thus, $\tilde{H}^{r}$ must have a Lax-Novikov integral  $\tilde{P}_{2g+2l+1}^{r}=P_{2g+2l+1}(\tilde{u}^{r}_{g,l}(x),\partial\sigma(H_{g,l}))$, such that 
$[\tilde{P}_{2g+2l+1}^{r},\tilde{H}^{r}_{g,l}]=-2i\frac{d}{dx}f_{g+l+1}(\tilde{u}^{r}(x),\partial\sigma(H_{g,l})).$
It is possible to connect the Lax operators through the intertwining operator (\ref{Xrl})
in the form
\be\label{xppx}
X(r)P_{2g+1}=\tilde{P}_{2g+1}^{r}X(r),
\ee
in order to prove this identity, it is  enough to note that $X(r)P_{2g+1}\left(X(r)P_{2g+1}\right)^\dag=\tilde{P}_{2g+1}^{r}X(r)\left(\tilde{P}_{2g+1}^{r}X(r)\right)^\dag$ and that any deformation of (\ref{xppx}) in the form $X(r)P_{2g+1}=\tilde{P}_{2g+1}^{r}X(r)+D(r)$, with $D(r)$ an intertwining operator between $H_{g,l}$ and $\tilde{H}^{r}_{g,l}$, is inconsistent for any order of $D(r)$.

The simplest example of auto Darboux transformation corresponds to chose  $\psi(r,x)$ as generator of the Darboux transformations to $u_{g,0}(x)$. The result of such transformation is 
\begin{equation}
\tilde{u}^{r}_{g,0}(x)=-2\frac{d^2}{dx^2}\ln(\theta(x{\bf v}+\tilde{{\boldsymbol \phi}}^{r},\tau))+\Lambda_0,
\end{equation}
a shape invariant transformation of $u_{g,0}(x)$ with ${\boldsymbol \phi}$ displaced to $\tilde{{\boldsymbol \phi}}^{r}={\boldsymbol \phi}+{\boldsymbol \alpha}(r)$.

{\bf Solitonic Darboux transformations}: these are Darboux transformations constructed using real states of $H_{g,l-1}$ of the form $\Psi[l]=\A_{l-1}\psi_{a_{l,1},a_{l,2}}(r_{l,1},r_{l,2},x)$, where $\A_{l-1}$, $\psi_{a_{l,1},a_{l,2}}(r_{l,1},r_{l,2},x)$, are defined in (\ref{psiaa}) and the paragraphs below it.

The intertwining operator 
\bea
A_l&=&\Psi[l]\frac{d}{dx}\frac{1}{\Psi[l]}\\ A_l^\dag &=& -\frac{1}{\Psi[l]}\frac{d}{dx}\Psi[l],\nonumber
\eea
intertwine
$ H_{g,l-1}=A_l^\dag A_l+z(r_{l,1})$ with $H_{g,l}=A_l A_l^\dag+z(r_{l,1})$, in the form
$A_lH_{g,l-1}=H_{g,l}A_l$ and $A_l^\dag H_{g,l}=H_{g,l-1}A_l^\dag$. The Darboux dres\-sing of the  Lax-Novikov integral $P_{2g+2(l-1)+1}$ of $H_{g,l-1}$ allows to find a Lax-Novikov integral $\hat{P}_{2g+2l+1}=A_lP_{2g+2(l-1)+1}A_l^\dag $ for $H_{g,l}$, such that $[P_{2g+2l+1},H_{g,l}]=0$. Since $P_{2g+2l+1}$ 
is two orders greater than $P_{2g+2(l-1)+1}$, it is possible to note that 
$u_{g,l}$ is a solution of an equation in s-KdVh of one order higher  than the initial solution $u_{g,l-1}$.

In \cite{PAM,5A}   the solitonic potentials and their res\-pective Lax-Novikov  integrals  for the free background and the Lam\'e periodic background have been studied in detail and   concrete norms for the cons\-truc\-tion of solitonic potentials have been established from the zeros theo\-rem. 
The simplest example of solitonic Darboux transformation is to use  $\psi_{a{1,1},a{1,2}}(r_{1,1},r_{1,2},x)$ as generator of a Darboux transformation to $u_{g,0}(x)$.
The result of such transformation is
\end{multicols}
\rule{91mm}{0.1mm}
\bea\label{ug1}
u_{g,1}(x)&=&-2\frac{d^2}{dx^2}\ln\left(a_{1,1}\theta(x{\bf v}+{\boldsymbol \phi}+{\boldsymbol \alpha}(r_{1,1}),\tau)\exp
\left(-i x \xi(r_{1,1})\right)\right.\nonumber\\& & \left.+a_{1,2}\theta(x{\bf v}+{\boldsymbol \phi}+{\boldsymbol \alpha}(r_{1,2}),\tau)\exp
\left(-i x \xi(r_{1,2})\right)\right)+\Lambda_0,
\eea
\rule{95mm}{0mm}\rule{91mm}{0.1mm}
\begin{multicols}{2}
\noindent the constants $\xi(r_{j,1})$ and $\xi(r_{j,2})$  present imaginary part different from zero for  $z(r_{j,1})$ in the forbidden bands of the spectrum of $H_{g,0}$; this is due to the exponential grow or decrease of non physical states.
Far from the center of the soliton defect, the solitonic Darboux transformation looks like an auto Darboux transformation because  only nonlinear phase shifts  remain  in the transformed potential. This is due to the asymptotic dominance of one of the exponential terms in (\ref{ug1}). Both asymptotic potentials correspond to potentials in Its-Matveev form with phase shifts $|\Delta{\boldsymbol \phi}_i|= |{\boldsymbol \alpha}_i(r_{1,2})-{\boldsymbol \alpha}_i(r_{1,1})|$, where $i=1,\ldots,g$ corresponds to the index of each vector component.
\end{multicols}
\begin{figure}[!h]
\centering
\includegraphics[scale=1]{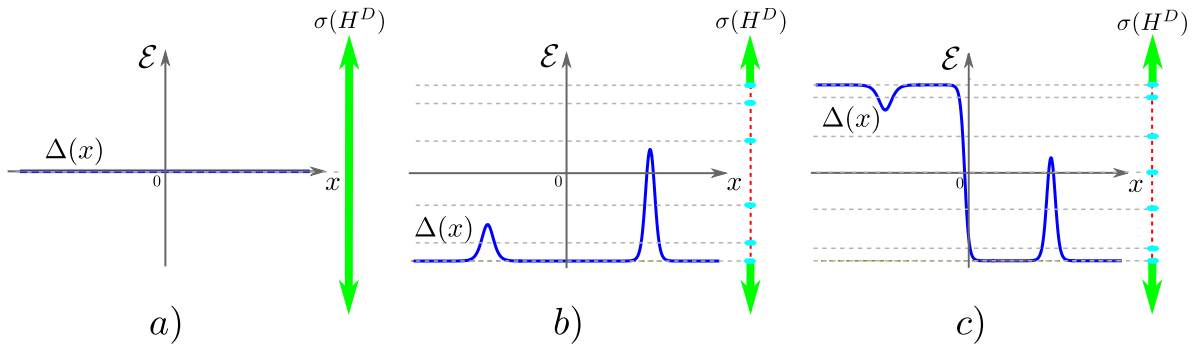}\\
\caption{Examples of stationary condensates of the GN model: solitonic defects on a homogeneous background. For each graphic are shown the shape of the condensate $\Delta(x)$, in blue continuous line, and its spectrum $\sigma(H^D)$, in green thick lines the continuous spectrum, in dashed red line the band gaps and in cyan dots the bound states and the band edges. Note that $-\Delta(x)$ is  a condensate solution of the GN model too, with identical spectrum. a) trivial zero solution, b) homogeneous background with two kink-antikink defects, c) kink on a homogeneous background with two kink-antikink defects.  }\label{fig1}
\end{figure}
\begin{figure}[!h]
	\centering
	\includegraphics[scale=1]{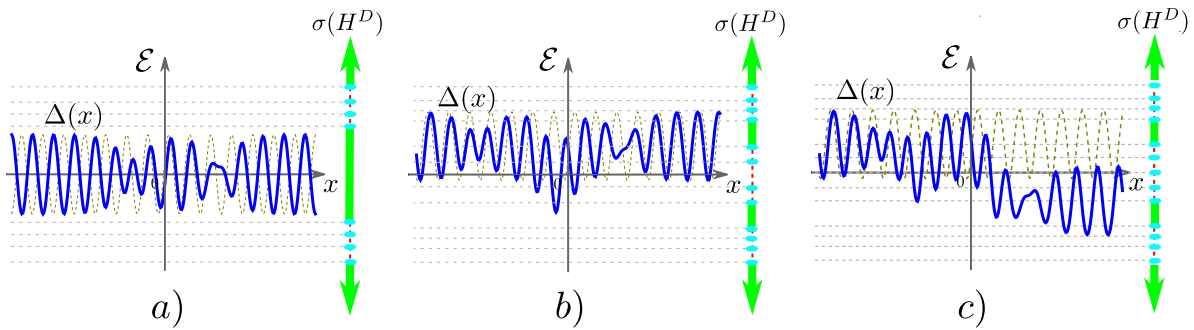}\\
	\caption{Examples of stationary condensates of GN model: solitonic defects on two- and three- gap  scalar Dirac potentials. It is used the same symbology as in Fig. (\ref{fig1}) but with a brown dashed thin   line added to show the shape of the finite-gap background,  a) two-gap condensate with two modulation shape solitonic defects that support each two bound states in the external gaps, one in each forbidden band, b) three-gap background with three solitonic defects, two of them as modulations and one kink-antikink defect that supports two bound states in the central band gap, c) kink version of previous condensate, its spectrum additionally contains   a zero energy state.   }\label{fig2}
\end{figure}
\begin{figure}[!h]
	\centering
	\includegraphics[scale=1]{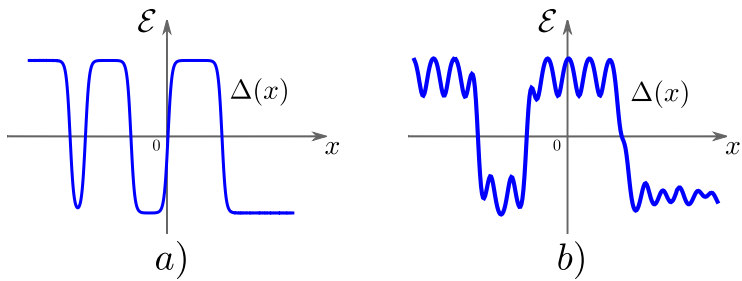}\\
	\caption{Kink defects look like a domain wall that pushes away two different phases of the condensate, while soliton defects with energies very close to zero allow to have two transitions between these two phases, this is due to the spontaneous appearance of a kink and an antikink. 
	The width of the kink-antikink increases while the energy of the defect approaches to zero, whilst in the zero limit one or both domain walls can disappear  in the spatial infinities.   }\label{fig3}
\end{figure}
\begin{multicols}{2}
\section{ Stationary condensates for the GN model}\label{S5}

It is possible to  differentiate between three types of stationary solutions for the GN model in dependence of their shapes and the spectrum of their respective Dirac operators: i) Kink finite-gap condensates: in this case, the condensate oscillates around zero and presents a central allowed band, see Fig. (\ref{fig1}{a}) and Fig. (\ref{fig2}{a}), ii) Kink-antikink finite-gap condensates: in this case the shape of the condensate oscillate around a constant different from zero and present a central gap, see Fig. (\ref{fig1}{b}) and Fig. (\ref{fig2}{b}), and  iii)  kink on Kink-antikink finite-gap condensates: these condensates are as\-so\-ciated to solitonic Darboux transformations, present a domain wall that divides two different Kink-antikink finite-gap phases. such domain wall is sup\-por\-ted by a single bound state of zero energy, see Fig. (\ref{fig1}{c}) and Fig. (\ref{fig2}{c}). Solitonic defects with Dirac energies  closer then zero allow to construct pairs of domain walls, see Fig. (\ref{fig3}).

{\bf Defects on kink-antikink finite-gap condensates} 

Through auto Darboux transformations
\be\label{vkak}
\Delta=-\frac{d}{dx}\ln(\A_l\psi(r_\star))
\ee
are constructed. The pair of Miura transformations that defines $\Delta$ in (\ref{vkak}) corresponds to $u_{g,l}- z_\star=\Delta^2-\frac{d}{dx}\Delta$ and $\tilde{u}^{r_\star}_{g,l}- z_\star=\Delta^2+\frac{d}{dx}\Delta$.

The described transformation allows to define the extended Schr\"odinger Hamiltonian $ \tilde{\mathcal H}={\rm diag}(H_{g,l},\tilde{H}^{r_\star}_{g,l})$, the fermionic integrals
\bea\label{Qdo1}
 \tilde{\mathcal{Q}}_{1}&=&\begin{pmatrix}0&X_{1,l}(r_\star)^\dag\\
X_{1,l}(r_\star)
&0\end{pmatrix}\\&=&\begin{pmatrix}0&-\frac{d}{dx}+\Delta\\
\frac{d}{dx}+\Delta
&0\end{pmatrix},\nonumber
\eea
and 
 $ \tilde{\mathcal{Q}}_2=i\sigma_3 \tilde{\mathcal{Q}}_1$, and also a
  Lax-Novikov integral in the form
 \begin{equation}\label{Pdo1}
 \tilde{\mathcal{P}}_{1}
 =
 \left(%
 \begin{array}{cc}
 P_{2g+2l+1}& 0 \\
 0 & \tilde{P}^{r_\star}_{2g+2l+1}\\
 \end{array}%
 \right),
 \end{equation}
that satisfies the following superalgebra
\be
[ \tilde{\mathcal H}, \tilde{\mathcal{Q}}_a]=0,\quad\{ \tilde{\mathcal{Q}}_a, \tilde{\mathcal{Q}}_b\}=2\delta_{ab}( \tilde{\mathcal H}- z_\star),
\ee
\begin{equation}\label{cons}
[ \tilde\cP_{1}, \tilde{\mathcal H}]={\bf 0}, \quad[ \tilde\cP_{1},\cQ_{a}]={\bf 0} ,
\end{equation}
the identities (\ref{cons}) correspond to the Lax pair formulation of the equations of both s-KdVh and s-mKdVh respectively.  Remember that the equations in the mKdVh are invariant under the change $\Delta\to-\Delta$ hence $\Delta= \frac{d}{dx}\ln(\A_l\psi(r_\star))$ also corresponds to a stationary GN condensate.

Additionally, it is possible to define $ \tilde\cP_{2}=\sigma_3 \tilde\cP_{1}$ which is another Lax-Novikov integral for  $ \tilde{\mathcal H}$, $[ \tilde\cP_{2}, \tilde{\mathcal H}]={\bf 0}$. The integral $\tilde\cP_{2}$ together with $\tilde\cQ_{a}$, $a=1,2$ define a pair of new fermionic integrals $[ \tilde\cP_{2},\tilde\cQ_{a}]$ that complete the $N=4$ fermionic integrals.

As a consequence of (\ref{cons}), the kink-antikink condensates $\Delta$ described in (\ref{vkak}) are solutions of $\hat f_{2g+2l+1}(\Delta)=0$ (see (\ref{hfl})),  which coefficients $c^D_k$ are given by the energies
$\partial\sigma( \tilde\cQ_1)$,
\end{multicols}
\rule{91mm}{0.1mm}
\bea
\sigma( \tilde\cQ_1)&=&(-\infty,-\sqrt{ E_0- z_\star}]\cup[-\sqrt{ E_1- z_\star},-\sqrt{ E_2- z_\star}]\cup\ldots\\
&\cup&[-\sqrt{ E_{2g-1}- z_\star},-\sqrt{ E_{2g}- z_\star}]\cup[\sqrt{ E_{2g}- z_\star},\sqrt{ E_{2g-1}- z_\star}]\cup\ldots\nonumber\\
&\cup&[\sqrt{ E_2- z_\star},\sqrt{ E_1- z_\star}]\cup[\sqrt{ E_0- z_\star},\infty)\nonumber\\
&\cup_{j=1}^l&\{-\sqrt{ z_j- z_\star},\sqrt{ z_j- z_\star}\},\nonumber
\eea
\rule{95mm}{0mm}\rule{91mm}{0.1mm}
\begin{multicols}{2}
\noindent where $\{E_0,\ldots,E_{2n}\}=\partial\sigma(H_{g,0})$ and $z_j=z(r_{j,1})= z(r_{j,2})$ are the solitons energies of the spectrum of $H_{g,l}$. 
Note that for a common base of eigenstates the eigenvalues $ \tilde \cE$ of $ \tilde\cQ_1$ and the eigenvalues $ \tilde E$ of $ \tilde\cH$ are related in the form $ \tilde \cE{}^2= \tilde E- z_\star$. 
This type of solutions, as seen in Fig. (\ref{fig2},b), correspond to self-consistent condensates $\Delta$ with a central energy band gap.

{\bf Defects on kink finite-gap condensates}

These are the limit of kink-antikink finite-gap condensates in which $\A_l\psi(r_\star)$ is the ground state of $H_{g,l}$ and it must also correspond to an edge of band, with energy $E_{2g}$. In this case $P_{2g+2l+1}$ and $X_{1,l}(r_\star)$ 
share $\A_l\psi(r_\star)$ as kernel, then there exist $S$ such that $P_{2g+2l+1}=S X_{1,l}(r_\star)$ and $\tilde P^{r_\star}_{2g+2l+1}=X_{1,l}(r_\star)^\dag S^\dag $. This produces an order reduction of the Lax Novikov central integral of the Dirac Hamiltonian operator $\tilde \cQ_1$, where $\tilde\cP_{1}=\tilde \cQ_1 \tilde \cS_1$,
\begin{equation}\label{TS}
\tilde \cS_1=\begin{pmatrix}0&S^\dag\\
S
&0\end{pmatrix}, \quad [\tilde\cS_1,\tilde\cQ_1]=0.
\end{equation}
 The identity (\ref{xppx}) defines $\tilde\cS_1$ and $\tilde\cQ_1$ as the Lax pair of 
 $\hat f_{2g+2l}(\Delta)=0$ in the s-mKdVh, with coefficients $c^D_k$ 
given by the boundaries of spectrum of $\tilde\cQ_1$,  $\partial\sigma(\tilde\cQ_1)$.
\end{multicols}
\rule{91mm}{0.1mm}
\bea
\sigma(\tilde\cQ_1)&=&(-\infty,-\sqrt{ E_0-E_{2g}}]\cup[-\sqrt{ E_1-E_{2g}},-\sqrt{ E_2-E_{2g}}]\cup\ldots\\
&\cup&[-\sqrt{ E_{2g-1}-E_{2g}},\sqrt{ E_{2g-1}-E_{2g}}]\cup\ldots\nonumber\\
&\cup&[\sqrt{ E_2-E_{2g}},\sqrt{ E_1-E_{2g}}]\cup[\sqrt{ E_0-E_{2g}},\infty)\nonumber\\
&\cup_{j=1}^{l}&\{-\sqrt{ z_j-E_{2g}},\sqrt{ z_j-E_{2g}}\}.\nonumber
\eea
\rule{95mm}{0mm}\rule{91mm}{0.1mm}
\begin{multicols}{2}
In this case $\tilde{\cQ}_a$, $\tilde \cS_1$ and $\tilde \cS_2=i\sigma_3\tilde \cS_1$ correspond to the four fermionic integrals of $\tilde{\cH}$.

{\bf Soliton defects plus Kink on kink-antikink finite-gap bakground}

Through solitonic Darboux transformations
\be\label{vk} 
\Delta=-\frac{d}{dx}\ln(\A_{l-1}\psi_{a_{l,1},a_{l,2}}(r_{l,1},r_{l,2},x)),
\ee 
are constructed. In this case, the pair of satisfied Miura transformations  are $u_{g,l-1}-z_l=\Delta^{2}-\frac{d}{dx}\Delta$ and $u_{g,l}-z_l=\Delta^{2}+\frac{d}{dx}\Delta$.

The extended Schr\"odinger Hamiltonian takes the form
$\mathcal H={\rm diag}(H_{g,l-1},H_{g,l})$
and its Lax-Novikov Integral can be written as
\be\label{Po1}
\mathcal{P}_{1}
=
\left(%
\begin{array}{cc}
(H_{g,l-1}-z_l)P_{2g+2(l-1)+1}& 0 \\
0 & P_{2g+2l+1}\\
\end{array}%
\right),
\ee
$[\cP_1,\cH]=0$. In the definition of  $\cP_1$ has been  introduced the term $(H_{g,l-1}-z_l)$ to have operators of the same order in its diagonal elements, which will be necessary in the next analysis.

In this case, it is easy to show that $\cP_1$ plays the role of  integral of motion for the  Dirac Hamiltonian operators 
\begin{equation}\label{Qo1}
\mathcal{Q}_{1}=\begin{pmatrix}0&A_l^\dag\\
A_l
&0\end{pmatrix}, \quad \mathcal{Q}_{2}=i\sigma_3\mathcal{Q}_{1},
\end{equation}
\be
[\cP_1,\mathcal{Q}_{a}]=0,
\ee
on the other hand in Dirac Hamiltonian frame: the kink nature of the superpotential in $\mathcal{Q}_{a}$ generates a spontaneous order reduction of the s-mKdVh Lax-Novikov operator $\cP_1$  because in this case it is possible to do the factorization $\cP_1=\cQ_1\cS_1$,
\bea\label{So1}
\mathcal{S}_1&=&
\begin{pmatrix}0&P_{2g+2(l-1)+1}A_l^\dag\\
A_lP_{2g+2(l-1)+1}
&0\end{pmatrix},\nonumber\\
\mathcal{S}_{2}&=&
i\sigma_3\mathcal{S}_{1},
\eea
from this point of view, it is allowed the order reduction $\cP_1\rightarrow\cS_1$. 
So taking  $\cQ_1$ as Dirac Hamiltonian, the irreducible integral of motion, in form of the s-mKdV
Lax-Novikov operator (\ref{mLN}), is not $\cP_1$ but rather $\cS_1$, 
\be
[\cQ_1,\cS_1]=0.
\ee
$\Delta$ are the solutions of $\hat f_{2g+2l}(\Delta)=0$ with coefficients $c^D_k$ 
given by the boundaries of spectrum of $\cQ_1$,  $\partial\sigma(\cQ_1)$.
\end{multicols}
\rule{91mm}{0.1mm}
\bea
\sigma(\cQ_1)&=&(-\infty,-\sqrt{ E_0-z_l}]\cup[-\sqrt{ E_1-z_l},-\sqrt{ E_2-z_l}]\cup\ldots\\
&\cup&[-\sqrt{ E_{2n-1}-z_l},-\sqrt{ E_{2n}-z_l}]\cup0\cup[\sqrt{ E_{2n}-z_l},\sqrt{ E_{2n-1}-z_l}]\cup\ldots\nonumber\\
&\cup&[\sqrt{ E_2-z_l},\sqrt{ E_1-z_l}]\cup[\sqrt{ E_0-z_l},\infty)\nonumber\\
&\cup_{j=1}^{l-1}&\{-\sqrt{ z_j-z_l},\sqrt{ z_j-z_l}\}.\nonumber
\eea
\rule{95mm}{0mm}\rule{91mm}{0.1mm}
\begin{multicols}{2}
The existence of a zero energy state is directly related to the kink or antikink nature of $ \Delta$.  

For a common base of states
the energies $\cE$ of $\breve\cQ=\cQ_1,\tilde\cQ_\star$ and the energies $E$ of $\breve\cH=\cH,\tilde\cH$ keep the relation $\cE{}^2=E-\breve z$, where $\breve z$ correspond to $z_l$ or $z_\star$ respectively.
The stationary condensates $\Delta$  have the characteristic property that their spectrums are always 
symmetrical with respect to $\cE=0$. 
 
\section{Solitonic defects and self-consistency }\label{S6}
The role of the  consistence equations (\ref{rn}) is to define the spectrum of fermion matter that composes each condensate.   
There exist infinite possibilities to fill the fermion states in the spectrum of a condensate $\Delta$. Among them, it is possible to differentiate between ground states and exciton solutions. 
In ground state configurations, a particle (hole) occupies (empties) a state only if those of smaller (greater) energy are also occupied (emptied), while exciton configurations allow discontinuities in the occupation of the states.

\begin{figure}[h!]
\centering
\includegraphics[scale=.5]{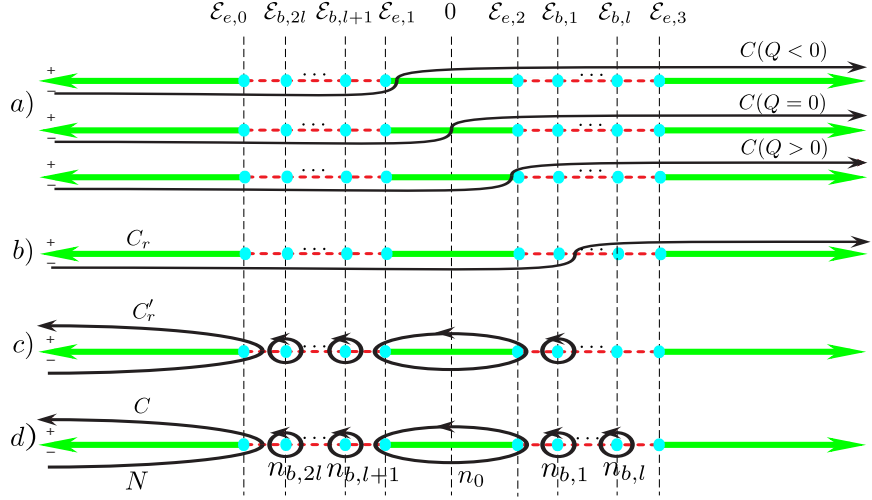}\\
\caption{Occupation of fermion matter. The occupation of the fermionic states by each flavor is characterized by the spectrum of the condensate $\Delta$. Such spectrum defines the poles and the branch cuts in the argument of the trace in the equation (\ref{cf}). Here, it is shown the meaning of the integration paths for ground state configurations: a) Examples of paths for ground states; the first path leaves holes in Dirac sea (antiparticles) so its charge $ Q $ will be "negative". The second case corresponds to  a full Dirac sea, where there are no particles nor antiparticles, thus, the total charge is zero. In the third case, there are only particles, so it has a "positive" charge.  Since the paths only cross the axis  once, these examples correspond to ground state type paths. The figure b) shows the path $C_r$ for $r$-th fermion flavor.  The path $C_r$ in b) is identical to the path $C_r'$ in c). $C'_r$ results from adding a  closed integration path    in the upper plane to $C_r$, such closed integration path covers the axis ${\rm Im}[z]=0$ with direction ${\rm Re}[z]=\infty\rightarrow {\rm Re}[z]=-\infty$ and whose integral, by means of the residue theorem, is	equal to zero. d) The superposition of the integration path of all flavors construct the path $C$. To simplify, the scattering zones, of particles and holes, are chosen empty. In other words, the lower allowed band is occupied by all the flavors, i.e. the path $C$ cover $N$ times the respective branch cut, and the upper allowed band is empty. On other hand, some flavors occupy the bound states, then the poles are turned by $C$ a number $n<N$ of times. Other assumption for simplicity is to choose all the states in the bands occupied by the same number of fermions. }\label{fig4}
\end{figure}

By using the identities (\ref{rn}) and (\ref{detN}), the consistency equation (\ref{consis}) for a condensate $\Delta$ with $n+1$ gaps and $2l$ defects, $n,l\in\mathbb{N}_0$, takes the form
\bea\label{cf}
i\frac{\Delta}{Ng^2}&=&{\rm tr}_{\cE}\left[ \frac{\cE^{n+2l}\Delta+\Sigma }{\sqrt{-\Pi}\prod_{k=1}^{l}(\cE^2-\cE_{b,k}^2 )} \right],\\
\Sigma&=&\sum_{j=1}^{\lfloor n/2\rfloor+l} \cE^{n+2l-2j}  \hat f_{2j}, \nonumber\\
\sqrt{-\Pi}&=&\sqrt{-\prod_{j=0}^{2n+1}(\cE-\cE_{e,j})},\nonumber
\eea
here $g^2$ is the coupling constant of the GN model\footnote{To avoid confusions, note that in  this section $g^2$ is not related to the genus of the Riemann theta function, for here $n$ takes this role.}, the constants $\cE_{e,j}$, for $j=0,1,\ldots,2n+1$, correspond to the energies of the edges of the bands of $H^D$ and $\cE_{b,k}=-\cE_{b,l+k}$, $0<\cE_{b,1}<\cE_{b,2}<\cdots<\cE_{b,l}$, to the energies of the bound states of $H^D$. Here ${\rm tr}_\cE$ is defined as 
\be\label{tra}
{\rm tr}_\cE\equiv \frac{1}{N}\sum_{r=1}^N\int_{C_r} \frac{d\cE}{2\pi}\equiv \frac{1}{N}\int_{C} \frac{d\cE}{2\pi},
\ee
where the integration paths $C_r$ in (\ref{tra}) depend on the spectrum occupied by the particles and anti-particles (holes) of flavor $r$, see Fig. (\ref{fig4}). Thus, the equation (\ref{cf}) correspond to a system of $\lfloor n/2\rfloor+l+1$ equations that defines the UV cut-off and the occupation of each state in the spectrum of $H^D$ by the different flavors.

 By observing that equation (\ref{cf}) must be fulfilled for any values of  $\hat{f}_{2j}=\hat{f}_{2j}(x)$, $j=1,\cdots, \lfloor n/2\rfloor+l$,  it is possible to reduce those equation in the following set of consistence equations: one equation for UV cut-off
\bea\label{uv}
\frac{i}{Ng^2}&=&{\rm tr}_\cE\left[\frac{\cE^n}{\sqrt{-\Pi}}\right],
\eea
$2\lfloor n/2\rfloor$ equations for the occupation of the bands 
\bea\label{cbands}
0&=&{\rm tr}_\cE\left[\frac{\cE^{n-j}}{\sqrt{-\Pi}}\right],
\eea
where $j=1,\cdots,2\lfloor n/2\rfloor$, and $l$ equations for the occupation of the bound states
\bea\label{cbound}
0&=&{\rm tr}_\cE\left[\frac{\cE^{n-2\lfloor n/2\rfloor}}{\sqrt{-\Pi}(\cE^2-\cE_{b,j}^2 )}\right],
\eea
where $j=1,\cdots,l$. The unknown variables of these equations are the occupation fractions $\nu(|\cE|)=\frac{n(-|\cE|)-n(|\cE|)}{N}$, $-1<\nu(|\cE|)<1$, where $n(\cE)$ is the number of flavors that occupy the eigenstate of allowed ener\-gy $\cE$.

\end{multicols}
\begin{multicols}{2}
 
It is interesting to note that the result of (\ref{cbound}) for each pair of defects  is independent of the other pairs. The occupation fraction $\nu_k=\nu(|\cE_{b,k}|)$ for the pair of bounded states of energies $|\cE_{b,k}|$ and $-|\cE_{b,k}|$  depends on the spectral information of the finite-gap background and  only on the modulo of the energy of such pair of defects, without taking into account the existence of the other pairs. This is a  characteristic already observed in the solitonic case in free massive background.

The consistence equations are the same for a kink condensate as for an identical condensate without the kink. In kink condensates there are $2l+1$ bound states, where one of such states has zero energy. In equation (\ref{cf}) this change  corresponds to  multiply numerator and denominator by $\mathcal{E}$, remaining unchanged the consistence equations. As consequence, the consistence equations (\ref{consis}) do not fill  the oc\-cu\-pation of the state of zero energy, and by definition $\nu(0)=0$.

In the following, the self-consistency of the condensates that present solitonic defects on one- and two-gap backgrounds  will be studied. For simplicity, a ground state occupation will by presented, in which the states of the lower allowed band are fully occupied, the states in the upper allowed band are completely empty and all the states in the allowed bands   have the same occupation fraction. 

One-gap solution corresponds to massive Dirac particle  $\Delta=\pm M$, $M>0$. In this case, Dirac o\-pe\-rator $\cQ$, see (\ref{Qdo1}), connects two copies of $H_{0,0}=-\frac{d^2}{dx^2}$. The eigenstates of $H_{0,0}$ correspond to 
$\psi(k)=e^{ikx}$, $H_{0,0}\psi(k)=E\psi(k)$, $E=k^2$, $k \in \R$ for physical states and  $k \in \I$ for non physical states. Through Miura-Darboux transformation, the massive condensate  rises from the state $\psi(i m)=e^{-mx}$, $m=\pm M$, being $\Delta=-\frac{d}{dx}\ln\psi(i m)$. Note that for $m=0$ the $0$-gap condensate is obtained through the physical edge of band state $\psi(0)=1$ of $H_{0,0}$.

The one-gap condensates with defects are obtained using non physical states of $H_{0,0}$ in the form
\begin{equation}
\Delta =-\frac{d}{dx}\ln\left(
\frac{\W\left(\breve\psi_{1},\cdots,\breve\psi_j,e^{-mx}\right)}
{\W\left(\breve\psi_{1},\cdots,\breve\psi_j\right)}\right)
\end{equation}
where $\breve\psi_{2\ell}=\sinh(\kappa_{2\ell}(x+\tau_{2\ell}))$ and $\breve\psi_{2\ell+1}=\cosh(\kappa_{2\ell+1}(x+\tau_{2\ell+1}))$, with $j$ and $\ell$ positive integers and $|\kappa_\ell|<|\kappa_{\ell-1}|<|m|$.
In this case, the spectrum of $\cQ$ cor\-res\-ponds to 
\bea
\sigma(\cQ)&=&(-\infty,-M]\cup_{i=1,\ldots,j}-\sqrt{M^2-\kappa^2_i}\nonumber\\
& &\cup_{i=1,\ldots,j}\sqrt{M^2-\kappa_i^2}\nonumber
\cup[M,\infty),
\eea
and the equations (\ref{cbound}) define the  occupation fraction $\nu_k$ of the bound states of the central band  in the form

\be \label{con1}
 |\cE_{b,j}|= M\sin\left(\frac{\pi\nu_j}{2} \right),\quad j=1,\cdots,l.
\ee
Note that for this type of condensates the occupation fractions are related by $\nu_1<\nu_2<\cdots<\nu_l $. This result allows solutions of ground state type, in which any particle of any flavor can exist only if there exists another of same flavor with the allowed energy immediately lower.  

Within the solutions on finite-gap background, the inhomogeneous simplest case (modulo spacial displacements) is the kink crystal background 
\bea\label{ck}
\Delta&=&-\left(\ln{\rm dn}(Mx,k)\right)'\\&=&Mk^2 \frac{\sn(Mx,k) \cn(Mx,k) }{\dn(Mx,k) },\nonumber
\eea
where $\sn(x,k)$, $\cn(x,k)$ and $\dn(x,k)$ are the Jacobi e\-llip\-tic functions with modular parameter $k$, $0<k<1$. The condensate (\ref{ck}) is a solution of the s-mKdVh, 
\be
0=\Delta ''(x)-2 \Delta (x)^3-2 \left(k^2-2\right) M^2 \Delta (x),\nonumber
\ee
this equation is also known as the non linear Schr\"odinger equation and corresponds to $\hat f_2(\Delta)=0$.

From now on, the following relationship between Jacobi's Theta function  and Riemann's Theta function will be used
\begin{equation}
\Theta(x|k)=\theta\left(x+\half,\tau\right),\quad \tau=i\frac{{\rm{\bf K}}'}{{\rm{\bf K}}} ,
\end{equation}
\be 
\dn(u,k)=\frac{\Theta(u+{\rm {\bf K}})}{\Theta(u)}
\frac{\Theta(0)}{\Theta({\rm {\bf K}})},
\ee
where ${\rm{\bf K}}(.)$   is the elliptic integral of the first kind,  ${\rm{\bf K}}={\rm{\bf K}}(k)$   is the complete elliptic integral of  the first kind and ${\rm{\bf K}}'={\rm{\bf K}}(k')$, $k'=\sqrt{1-k^2}$. 

From the supersymmetric point of view, the kink crystal condensate  connects two Lam\'e one-gap Schr\"odinger potentials displaced in half a period:  $u_{1,0}=M^2k^2(2\sn{(Mx,k)^2}-1)$ and $\tilde u^{(0,0)}_{1,0}=M^2k^2(2\sn{(Mx+{\rm{\bf K}}(k),k)^2}-1)$. The Darboux transformation that generates this superpotential is constructed using the ground state $\psi={\rm dn}(Mx,k)$ of the one-gap Schr\"odinger system $H_{1,0}$. Such Darboux transformation allows to construct a kink two-gap condensate, with two band gaps separating a central band \cite{PlyANie}.
 
 This condensate has a period of $2 {{\rm{\bf K}}}(k) $, 
defined in terms of the complete elliptic integral of first kind  ${{\rm{\bf K}}}(k) $. 

 The spectrum of kink crystal (\ref{ck}) is given by $\sigma(H^D)= (-\infty, -M] \cup [-Mk',Mk'] \cup [M, \infty) \ $, where the energy band gaps correspond to $-M< \cE < -Mk' $ and $ Mk' < \cE < M $.

The eigenstates of $H_{1,0}$, $H_{1,0}\psi_\pm^{\alpha}=E(\alpha)\psi_\pm^{\alpha}$ are 
\begin{equation}\label{psi+-}
\psi_\pm^{\alpha}\left(x\right)=
\frac{{\rm H}\left(Mx\pm \alpha\right)}{\Theta\left(Mx\right)}
\exp\left[
\mp Mx {\rm Z}\left(\alpha\right)\right]\,,
\end{equation}
where $E(\alpha)=M^2{\rm dn}^2(\alpha,k)$ and  ${\rm Z}$ and ${\rm H}$ are the Jacobi Zeta and Eta functions respectively,
\begin{equation}\label{ZTheta}
{\rm Z}(u|k)=\frac{d}{du}\ln \Theta(u),
\end{equation}
\begin{equation}\label{Eta}
{\rm H}(u)=-iq^{1/4}\exp\left(i\frac{\pi u}{2{\rm{\bf K}}}\right)
\Theta(u+i{\rm{\bf K}}')\,.
\end{equation}
\begin{equation}\label{q}
q=\exp(i\pi  \tau)\,.
\end{equation} 
For more details about these functions and the spectral properties such as quasi-momentum, see \cite{PAM}. The functions $\Psi_\pm^{\alpha}$ are parametrized in terms of the parameter $\alpha$, which lies in a rectangular domain with vertices $\alpha \in \{0,{\rm{\bf K}},{\rm{\bf K}}+i {\rm{\bf K}}', i{\rm{\bf K}}'\}$.

By using chains of Darboux transformations, it is possible to construct an infinite set of symmetric condensates  with one central allowed band in the form
	\be\label{D2l}
	\Delta(x)=-\frac{d}{dx}\ln\left(\frac{W(\phi(1),\ldots,\phi(2\ell),\dn(M x,k))}
	{W(\phi(1),\ldots,\phi(2\ell))}\right),
	\ee
\begin{equation}\label{phi+}
\phi(2j+1)\equiv C_{2j+1}\psi^{\alpha_{2j+1}}_+(x)+
\frac{1}{C_{2j+1}}\psi^{\alpha_{2j+1}}_-(x),	
\end{equation}
\begin{equation}\label{phi-}
\phi(2j)\equiv C_{2j}\psi^{\alpha_{2j}}_+(x)-
\frac{1}{C_{2j}}\psi^{\alpha_{2j}}_-(x),	
\end{equation}
\begin{equation}\label{beta+beta-}
0<\alpha_1<\alpha_2<...<\alpha_{2\ell}<{\rm{\bf K}}
\end{equation}

The spectrum of these condensates is defined by 
\bea
\sigma(H^D)&=& (-\infty, -M]\cup_{i=1}^{2\ell}\{-M\dn(\alpha_i,k)\}\nonumber\\
 & &\cup [-Mk',Mk']\nonumber\\
 & & \cup_{i=1}^{2\ell}\{M\dn(\alpha_i,k)\}\cup [M, \infty).\nonumber
\eea
It is possible to obtain potentials with $2(2\ell-1)$ bound states, one way to obtain these is to choose a parameter $C_j$ and to take either the limit $C_j\rightarrow 0$ or $\infty$. With this method,  one of the solitons disappears in some of the two spatial infinities, $x\rightarrow\pm\infty$. As a consequence of the loss of two bound states, the characteristic equation in s-mKdVh  of the initial condensates  will fall by two orders. In summary, through this method stationary condensates $\Delta$ are constructed, whose spectra are symmetrical with respect to $ \cE = 0 $; they also have a finite number of bands and any number of bound states in them.

 Due to the different occupation number of each sector of the spectrum, the occupation fraction $\nu_0$ is introduced for the occupation of the states in the central band and $\nu_{j}$, $j=1,\cdots,l$, for the bound states.

Because of the symmetries of the spectrum, the oc\-cu\-pation constant $\nu_0$ is undefined as in kink case. Thus, the consistency equation (\ref{cbound}) for the condensate (\ref{D2l}) takes the form
\be
\nu_l=\frac{2}{\pi }\tan^{-1}\left(\frac{\sqrt{\cE_{b,l}^2-k'{}^2 M^2}}{\sqrt{M^2-\cE_{b,l}^2}}\right),
\ee
the result shows how the effective occupation fraction of the defects  relies on  their energy and  width of the central band (or the same, the modular parameter $ k$).  The inverse interpretation defines the energy of the defect in function of the effective occupation fraction and the width of the central band 
\be
|\cE_{b,l}|= M \cos \left(\frac{\pi  \nu_1 }{2}\right) \sqrt{k'^2+\tan ^2\left(\frac{\pi  \nu_1 }{2}\right)},
\ee
it is interesting to notice that in limit $k'\rightarrow 0$  known results of the solitonic case in free massive background are obtained (\ref{con1}). In this limit $k\rightarrow 1$, $ {{\rm{\bf K}}} \rightarrow \infty$ and the kink crystal
(\ref{ck}) is  reduced to the Callan-Coleman-Gross-Zee kink $\Delta= M\tanh (Mx)$  \cite{DHN}.

For more complex condensates, the consistence equations depend on hyper-elliptic integrals of higger order.

\section*{Discussion and outlook}

By using an exotic supersymmetry between finite-gap systems with defects, the set of  analytical stationary solutions for the GN model has been constructed, obser\-ving the existence of inhomogeneous and non-periodic condensates  with band structures and a finite number of bound states.

The Darboux transformation has allowed  to recursively construct  infinite families of exactly solvable Scr\"odinger systems from a finite-gap potential given by the Its-Matveev formula.

The process of constructing solitary defects on  finite-gap  Schr\"odinger potentials has generated scalar Dirac potentials ( or Bogoliubov-de Gennes self-consistent condensates) that present solitary defects on  finite-gap backgrounds. Each one of these Dirac systems exhibit an irreducible integral of motion  correspon\-ding to a Lax operator of the s-mKdVh.

The Darboux dressing of the Lax operator of finite-gap systems has allowed to find the self-consistency equations  for all the stationary condensates, having as main characteristic the independence of the consistence equations for each pair of bound states of opposite energy,  each one depending only on finite-gap background data and the modulus of its characteristic energy.

On dependence of their shape and spectrum, the set of self-consistent stationary condensates presented here can be separated into three groups. These condensates necessarily have the form of: i) kink finite-gap condensates with solitonic defects, which oscillate around zero and present a central allowed band, ii) kink-antikink finite-gap condensates with solitonic defects, which oscillate around a constant different from zero and present a central forbi\-dden band, and iii) kink domain wall condensates on kink-antikink finite-gap bakground with solitonic defects, which looks like  a kink (antikink) domain wall that pushes away two kink-antikink finite-gap phases and supports one bound state of zero energy in the central forbidden band; the two kink-antikink finite-gap phases oscillate around opposite constants.

In what follows, some interesting related  study problems will be discussed.
Important aspects to be studied about the condensates presented here are their mass spectrum  and their stability. In the direction of the first point, the study of the Green function and the density of states of the respective Schr\"odinger and Dirac equations for finite-gap systems with defects, are within the research interests of the author. About the stability of these condensates and their decay channels \cite{JF}: the possible decays of allowed bands into bound states or allowed bands into another allowed bands are mysteries yet to be revealed. 

The thermodynamic role that these condensates could occupy in GN model is another interesting point of study, being the most direct approach the Ginzburg-Landau expansion of the thermodynamic grand potential \cite{ThiesLame,BDT,FCGDMP}.

On the other hand, as well as s-mKdVh is related to the GN model, ZS-AKNS is related to the chiral GN model or Nambu-Jona-Lasinio model in 1+1D. There exists a generalization of Darboux transformation for Dirac operators in 1 + 1D. Particularly, such transformation allows the construction of pseudo-scalar Dirac potentials in form of soliton defects on finite-gap backgrounds with spectrum not necessarily symmetric. Although these potentials also present a Lax-Novikov integral, they are not solutions of the s-mKdVh, rather of the ZS-AKNS hierarchy of equations \cite{ZS, AKNS}. In general, these potentials cor\-res\-pond to  complex pseudo-scalar potentials with asym\-me\-tric spectra, any number of bound states and a finite number of energy band gaps. Due to the existence of the Lax-Novikov integral and an adequate Darboux transformation, it is possible to find a  representation for a nonlinear $ N = 4 $ supersymmetry   for extended Dirac Hamiltonians.  

The supersymmetric method has proved to be useful for solving problems of nonlinear interaction between bosons \cite{KayMos,PAM2} and problems of nonlinear interaction between fermions, herein studied. In this direction, it is interesting to search for uses of the supersymmetric methods for the study of coupled systems between bosons and fermions.

\vskip0.2cm
\noindent \textbf{Acknowledgements}
The author is thankful to Dr. Mikhail Plyushchay and Dr. Mokhtar Hassaine for critical
comments and to Roberto Mu\~noz, Claudia Baeza and Guillermo Guerra for their kind help and hospitality. The research was partially supported from FONDECYT through grant 3170474.

\end{multicols}
 
\end{document}